\documentclass[reprint,amsmath,amssymb,aps]{revtex4-1}

\usepackage[T1]{fontenc}
\usepackage{nicefrac}
\usepackage{color}
\usepackage{hyperref}
\usepackage{amsmath}
\usepackage{amssymb}

\usepackage{graphicx}

\newcommand{\psec}[1]{\subsection{#1}}

\newcommand{\rmd}{{\rm d}}

\newcommand{\Om}{\Omega_m}
\newcommand{\dc}{\delta_c}
\newcommand{\denv}{\delta_{\rm env}}
\newcommand{\denvl}{\delta_{\rm L, env}}
\newcommand{\denvnl}{\delta_{\rm NL, env}}

\begin{document}

\title{Consistency of the local Hubble constant with the cosmic microwave background}

\author{Lucas Lombriser}

\affiliation{D\'{e}partement de Physique Th\'{e}orique, Universit\'{e} de Gen\`{e}ve, \\ 24 quai Ernest Ansermet, 1211 Gen\`{e}ve 4, Switzerland}

\date{\today}

\begin{abstract}
A significant tension has become manifest between the current expansion rate of our Universe measured from the cosmic microwave background by the \emph{Planck} satellite and from local distance probes, which has prompted for interpretations of that as evidence of new physics.
Within conventional cosmology a likely source of this discrepancy is identified here as a matter density fluctuation around the cosmic average of the 40~Mpc environment in which the calibration of Supernovae Type~Ia separations with Cepheids and nearby absolute distance anchors is performed.
Inhomogeneities on this scale easily reach 40\% and more.
In that context, the discrepant expansion rates serve as evidence of residing in an underdense region of $\denv\approx-0.5\pm0.1$.
The probability for finding this local expansion rate given the \emph{Planck} data lies at the 95\% confidence level.
Likewise, a hypothetical equivalent local data set with mean expansion rate equal to that of \emph{Planck}, while statistically favoured, would not gain strong preference over the actual data in the respective Bayes factor.
These results therefore suggest borderline consistency between the local and \emph{Planck} measurements of the Hubble constant.
Generally accounting for the environmental uncertainty, the local measurement may be reinterpreted as a constraint on the cosmological Hubble constant of $H_0=74.7^{+5.8}_{-4.2}$~km/s/Mpc.
The current simplified analysis may be augmented with the employment of the full available data sets, an impact study for the immediate $\lesssim10$~Mpc environment of the distance anchors, more prone to inhomogeneities, as well as expansion rates measured by quasar lensing, gravitational waves, currently limited to the same 40~Mpc region, and local galaxy distributions.
\end{abstract}

\maketitle

\renewcommand\thesubsection{\arabic{subsection}}

\psec{Introduction}
%
Determining the expansion history of our Universe is paramount to resolving its energetic composition or testing key ingredients of standard cosmology such as the validity of General Relativity on large scales and the cosmic principle.
Measurements of the current expansion rate reveal a significant tension between the constraints inferred from the cosmic microwave background (CMB)~\cite{Aghanim:2018eyx} and those found from separations to Supernovae Type~Ia (SN~Ia) using absolute distance anchors from nearby Cepheids, masers, parallaxes, and ecliptic binaries~\cite{Riess:2009pu,Riess:2011yx,Efstathiou:2013via,Riess:2016jrr,Riess:2019cxk}.
The $4.4\sigma$~\cite{Riess:2019cxk} discrepancy between these probes of the early and late Universe has hence spurred much speculation for its interpretation as a signature of new physics (see for example Ref.~\cite{Freedman:2017yms}).
Independent measurements of the Hubble constant such as from quasar lensing~\cite{Wong:2019kwg} further increase this tension, and important insights are expected from Standard Sirens, which however currently do not give a clear preference for either value of the expansion rate~\cite{Abbott:2017xzu}.

This \emph{Letter} argues that the local measurement of the Hubble constant is likely impacted by the matter density fluctuation around the cosmic average of the 40~Mpc environment surrounding us.
Such an inhomogeneity is shown here to affect the calibration of SN~Ia distances with the Cepheids and the anchors, performed in the same volume.
The scale is set by the furthest host galaxy of the sample~\cite{Riess:2016jrr}.
Previous studies (see Ref.~\cite{Kenworthy:2019qwq,Wu:2017fpr} and references therein) have focused on underdense environments at scales of several hundreds of Mpc, where inhomogeneities are however strongly constrained in amplitude.
Density fluctuations on a 40~Mpc scale in contrast may easily reach 40\% and more.

This work presents an estimation of the likelihood of residing in a sufficiently underdense 40~Mpc environment to give rise to the observed discrepancy between the measured local and cosmological Hubble constants.
The numerical computations performed for this purpose will assume \emph{Planck}~\cite{Aghanim:2018eyx} mean cosmological parameters and standard deviations, specifically the total matter density parameter $\Om=0.315\pm0.007$, the Hubble constant $H_0=(67.4\pm0.5)$~km/s/Mpc, and the matter fluctuation amplitude $\sigma_8=0.811\pm0.006$.
The local expansion rate is given by $\hat{H}_0=(74.03\pm1.42)$~km/s/Mpc~\cite{Riess:2019cxk}, where hats will denote local quantities for clarity.
The speed of light in vacuum is set to $c=1$.

\psec{Calibration of SN~Ia distances}
%
A measurement of the local Hubble constant $\hat{H}_0$ was conducted in Ref.~\cite{Riess:2009pu}, employing a distance ladder that is based on precise maser separations to NGC~4258 to calibrate Cepheid distances and from that separations to SN~Ia using suitable host galaxies.
Improved measurements adopting this technique were performed in Refs.~\cite{Riess:2011yx,Efstathiou:2013via,Riess:2016jrr,Riess:2019cxk}, including more hosts for the Cepheid-calibrated SN~Ia separations as well as additional independent calibrations of the absolute Cepheid distances from Milky Way parallaxes and from Cepheids and detached eclipsing binary separations in the Large Magellanic Cloud.
The combination of these distance probes currently yields a measurement of $\hat{H}_0=(74.03\pm1.42)$~km/s/Mpc~\cite{Riess:2019cxk}, which is in 4.4$\sigma$ tension with the CMB measurement of \emph{Planck}.

More specifically, the measurement of the local Hubble constant is obtained from the relation~\cite{Riess:2009pu} (App.~\ref{sec:H0measurement})
\begin{equation}
 \log \hat{H}_0 = \frac{1}{5}(m^0_{x,{\rm N4258}}-\mu_{0,{\rm N4258}}) + a_x + 5  - \log \hat{d}_1 \,,
 \label{eq:localH0}
\end{equation}
where $m^0_{x,{\rm N4258}}$ is the expected peak magnitude of a SN~Ia in NGC~4258
and $\mu_{0,{\rm N4258}}$ denotes the independent precise distance modulus from its masers, which relates to the luminosity distance $\hat{d}_{L,{\rm N4258}}$ as $\mu=5\log (\hat{d}_L/\hat{d}_1) + 25$.
The peak magnitude $m^0_{x,{\rm N4258}}$ is found from a fit to the joint Cepheid/SN~Ia data, where the $j$-th measured Cepheid magnitude in the $i$-th host $m_{H,i,j}^W$ provides the difference in the distance moduli $\mu_{0,i}-\mu_{0,{\rm N4258}}$, which corresponds to the difference $m_{x,i}^0 - m^0_{x,{\rm N4528}}$ for the SN~Ia magnitudes $m_{x,i}^0$~\cite{Riess:2016jrr}.
The term $a_x$ is the intercept of the magnitude-redshift relation for the SN~Ia, given by $a_x = \log (H_0 d_L) - 0.2 m_x^0$, which importantly is independent of an absolute separation scale.
The luminosity distance $d_L$ in $a_x$ is fit to a large SN~Ia sample covering the redshifts $0.023<z<0.15$, $\sim$(100--650)~Mpc, and the individual $m_x^0$ are determined by a light-curve fitter.

Note that the distance normalisation $\hat{d}_1$ in Eq.~(\ref{eq:localH0}) has been kept explicit here and refers to a 1~Mpc absolute distance in the local reference frame.
For $\hat{d}_1$ to simply cancel it must be given in the same reference frame as $\hat{d}_{L,{\rm N4258}}$.
For a homogeneous universe, the cosmological ruler equals the local one, $d_1=\hat{d}_1$, such that Eq.~(\ref{eq:localH0}) also implies a measurement of the cosmological $H_0$.
In the case of a local inhomogeneity, however, the local and cosmological metrics are related by a conformal transformation $\hat{g}_{\mu\nu}= (\hat{a}/a)^2 g_{\mu\nu}$ for the respective scale factors $\hat{a}$, $a$ (App.~\ref{sec:frames}).
Unlike for coordinate transformations, conformal transformations do not preserve physical distances, and it is easy to show that $\hat{H}_0/H_0 = d_L/\hat{d}_L$.
Hence, Eq.~(\ref{eq:localH0}) implies a measurement of the expansion rate of the local environment, defining the frame in which $\hat{d}_{L,{\rm N4258}}$ is measured.
Given $\hat{d}_{L,{\rm N4258}}$ but not $d_{L,{\rm N4258}}$, one also arrives at this conclusion if expressing Eq.~(\ref{eq:localH0}) in terms of $H_0$ and $d_1$ instead.
The use of an absolute distance anchor in the cosmological frame in contrast would provide a measurement of $H_0$, consistent with the findings of Ref.~\cite{Macaulay:2018fxi}.

In principle the absolute distance scale only enters through NGC~4258 at 7.5~Mpc and other more nearby anchors.
One may therefore consider the relevant environment as set by that scale.
However, the calibration of the Cepheid magnitudes $m_{H,i,j}^W$ from which the scale in $\mu_{0,i}$ is inferred is performed with an involved fitting procedure of Cepheid parameters to the full sample of the joint Cepheid/SN~Ia hosts at $\lesssim40$~Mpc.
The environment for the absolute distance anchors is therefore conservatively set here to $\lesssim40$~Mpc.
The effect of an environmental density fluctuation for the absolute distance anchors at $\lesssim10$~Mpc lies beyond the scope of this work and is left for future study.
Whereas matter density fluctuations on the scales of (100--650)~Mpc of the high-$z$ SN~Ia samples are limited to standard deviations of $\lesssim$(5-20)\% around the mean, these reach about 40\% on scales of 40~Mpc and may hence impact the locally measured expansion rate.

\psec{Expansion rate of the local environment}
%
The luminosity distances $d_L$ entering Eq.~\eqref{eq:localH0} are related to the background expansion history $H(z)$ as $d_L(z)=(1+z)\int_0^{z} \rmd\tilde{z}/H(\tilde{z})$, where a spatially flat statistically homogeneous and isotropic universe with standard cosmological components of baryons, cold dark matter, and a cosmological constant $\Lambda$ is assumed throughout.
Radiation will be neglected for the late-time universe of interest.
From the resulting time-time component of the Einstein field equations with the Friedmann-Lema\^itre-Robertson-Walker (FLRW) metric, it follows that
\begin{equation}
 H^2\equiv \left(\frac{\rmd \ln a}{\rmd t}\right)^2 = \frac{8\pi G_N}{3}\bar{\rho}_m + \frac{\Lambda}{3} \,,
\end{equation}
where $a(t)=(1+z)^{-1}$ is the scale factor of the FLRW metric in terms of cosmological time $t$ normalised today, $G_N$ denotes Newton's gravitational constant, $\bar{\rho}_m$ represents the total matter density in the cosmological background, and $H_0\equiv H(z=0)$.

The local environment may be treated as a separate FLRW universe embedded in the larger Cosmos~\cite{Gunn:1972sv}.
This ansatz is well motivated~\cite{Dai:2015jaa} since with $r_{vir}\approx0.2$~Mpc the virial radii of the Milky Way and NGC~4258 are much smaller than the wavelength of the 40~Mpc density fluctuation.
Energy conservation in this environment with respect to cosmological time $t$ implies a local expansion rate of
\begin{equation}
 \hat{H} = -\frac{1}{3} \frac{ \rmd \ln \hat{\rho}_m}{\rmd t} \,, \label{eq:localexpansion}
\end{equation}
where $\hat{\rho}_m$ is the average environmental matter density, which relates to that of the cosmological background as $\hat{\rho}_m \equiv \bar{\rho}_m (1+\denv)$.
This defines the local environmental matter density fluctuation $\denv$.

The evolution of $\denv$ is described by the solution of Eq.~\eqref{eq:y} and can be cast as a function of $\hat{H}_0/H_0$.
Thus, the measured discrepancy of $\hat{H}_0/H_0\simeq1.1$ between the Hubble constant obtained from \emph{Planck} and from the local probes may be interpreted as a measurement of the local environmental inhomogeneity, averaged over a $40$~Mpc scale.
For simplicity keeping cosmological parameters fixed, except for the expansion rates, and casting data and probabilities into Gaussians, one infers from the measurements of $\hat{H}_0$ and $H_0$ a local density
with mean and standard deviation of approximately $\delta_{{\rm env},0} \approx -0.5\pm0.1$,
where errors have been added in quadrature and $\Delta \delta_{{\rm env},0} = \left| \frac{\partial \delta_{{\rm env},0}}{\partial (\hat{H}_0/H_0)} \Delta (\hat{H}_0/H_0) \right|$.

In the following an estimate for the likelihood of residing in such a local environment will be provided, noting that the required underdensity lies in the nonlinear regime of structure formation.

\begin{figure*}
\centering
 \resizebox{0.45\textwidth}{!}{
 \includegraphics{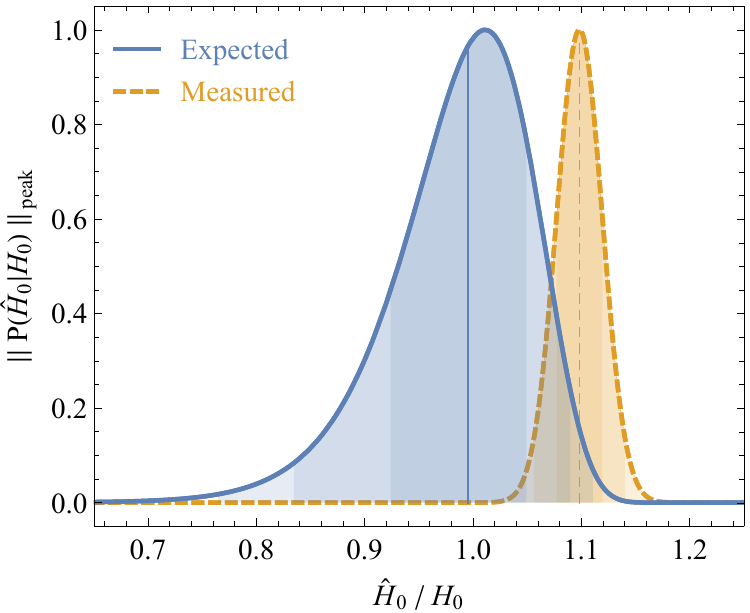}
 }
\resizebox{0.45\textwidth}{!}{
 \includegraphics{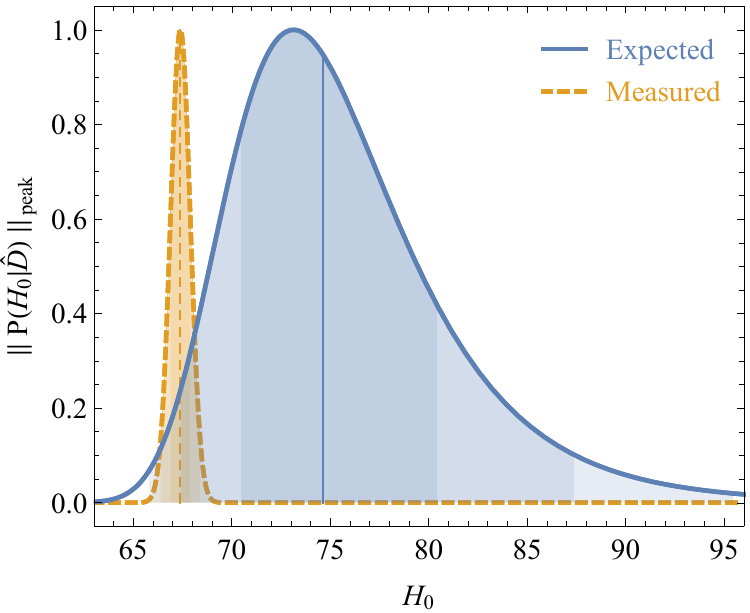}
 }
\caption{
Peak-normalised probabilities for the expected expansion rate $\hat{H}_0$ in our local 40~Mpc environment given a cosmological value $H_0$ (\emph{left panel}) and for a cosmological $H_0$ given the local data $\hat{D}$~\cite{Riess:2019cxk} cast into a Gaussian (\emph{right panel}).
Shaded regions illustrate 68\%, 95\%, and 99\% confidence levels around the median.
The Gaussianised local (\emph{left panel}) and \emph{Planck}~\cite{Aghanim:2018eyx} (\emph{right panel}) measurements are shown for comparison (dashed curves), adopting the mean $H_0$ of \emph{Planck} for $\hat{H}_0/H_0$.
The expansion rates marginally agree at the 95\% confidence level, and there would be no strong evidence in the respective Bayes factor for a hypothetical equivalent local measurement with equal mean to that of \emph{Planck} compared to the actual data.
}
\label{fig:results}
\end{figure*}

\psec{Nonlinear evolution of matter densities}
%
For simplicity, the environmental density fluctuation will be treated as a spherically symmetric top hat.
From energy-momentum conservation, $\nabla^{\mu}T_{\mu\nu}\rvert_{\rm env}=0$, one derives the nonlinear evolution equation~\cite{Gunn:1972sv,1980lssu.book.....P}
\begin{equation}
 y'' + \left( 2 + \frac{H'}{H} \right) y' + \frac{1}{2} \Omega_{\rm m}(a) \left( y^{-3} - 1 \right) y = 0 \label{eq:y}
\end{equation}
for the dimensionless physical top-hat radius $y=(\hat{\rho}_{\rm m}/\bar{\rho}_{\rm m})^{-1/3}$, where primes denote derivatives with respect to $\ln a$ and $\Omega_{\rm m}(a) \equiv 8\pi G_N \bar{\rho}_{\rm m}/(3H^2)$.
The evolution of $y$ can be determined setting initial conditions in the matter-dominated regime $a_i\ll1$, where $y_i \equiv y(a_i) = 1 - \delta_{{\rm env},i}/3$ and $y_i' = - \delta_{{\rm env},i}/3$ for an initial top-hat fluctuation $\delta_{{\rm env},i}$.
For an environment that undergoes spherical collapse today, one can define the current linear spherical collapse density, which for the \emph{Planck} cosmological parameters is given by $\dc=1.678$.
Note that for small fluctuations $y\approx1$, Eq.~\eqref{eq:y} can be linearised, yielding the usual evolution equation for the linear growth function of cosmological structure $(\delta_{\rm L}/\delta_{{\rm L},i})(a)$.

\psec{Distribution of environmental densities}
%
The probability for a linear environmental density fluctuation $\denvl$, defined by the Eulerian (physical) radius $\zeta = 40$~Mpc, can be estimated using excursion set theory, where the resulting distribution is approximately described by the expression~\cite{Lam:2007qw}
\begin{eqnarray}
 P_{\zeta}(\denvl) & = & \frac{\beta^{\omega/2}}{\sqrt{2\pi}} \exp \left[ -\frac{\beta^{\omega}}{2} \frac{\denvl^2}{(1 - \denvl/\dc)^{\omega}} \right] \nonumber\\
 & & \times \left[ 1 + (\omega - 1)\frac{\denvl}{\dc} \right] \left( 1 - \frac{\denvl}{\dc} \right)^{-\omega/2-1} \label{eq:eulerian}
\end{eqnarray}
with $\beta = (\zeta/8)^{3/\dc} / \sigma_8^{2/\omega}$, $\omega = \dc \gamma$, and $\gamma = - \frac{\rmd \ln S_{\xi}}{\rmd \ln M_{\rm env}} = 1+\tilde{n}_{\rm s}/3$.
The Lagrangian (or initial comoving) radius is set by $\xi=8$~Mpc/h such that $S_{\xi}=\sigma_8^2$.
The slope of the linear matter power spectrum $P_{\rm L}(k)$ on large scales at an initial time $a_{\rm i}\ll1$ in the matter era after the turn over is set to $\tilde{n}_s=-1.6$.

The probability for the nonlinear density fluctuations $P_{\zeta}(\denvnl)$ can be determined from employing the linear growth function $(\delta_{\rm L}/\delta_{{\rm L},i})(a)$ to evolve the range of $\denvl$ back to an initial time $a_i\ll1$ during the matter-dominated epoch, which is well described by linear theory, and then evolve the initial fluctuations $\delta_i$ forward to the present time with the nonlinear solutions of Eq.~\eqref{eq:y}.
The resulting probability distribution $P(\hat{H}_0\vert H_0)$ for a local Hubble constant $\hat{H}_0$ given $H_0$ in the cosmological background inferred from $P_{\zeta}(\denvnl)$ using Eq.~\eqref{eq:localexpansion} is shown in Fig.~\ref{fig:results} and is in good agreement with the distribution measured from $N$-body simulations in Ref.~\cite{Wojtak:2013gda} (also see Ref.~\cite{1992AJ....103.1427T}), where with $0.88\lesssim\hat{H}_0/H_0\lesssim1.15$ at $2\sigma$ the simulations allow for even larger, less conservative, fluctuations at the relevant upper bound.

\psec{Consistency of local $\hat{H}_0$ with CMB $H_0$}
%
The level of agreement or disagreement between the local and CMB measurements of the Hubble constant can now be quantified.
For a simple comparison, the local $\hat{H}_0$ measurement of Ref.~\cite{Riess:2019cxk} is shown alongside with $P(\hat{H}_0\vert H_0)$ in Fig.~\ref{fig:results}, where both distributions are normalised by their peak amplitudes and $\hat{H}_0$ has been divided by the mean \emph{Planck} value of $H_0$ for illustration.
The local measurement is found here to lie at the 95\% confidence level of $P(\hat{H}_0\vert H_0)$, suggesting borderline consistency with the \emph{Planck} value.
However, for a better assessment of the consistency between the two measurements, one may wish to directly estimate the probability of one measurement given the other.
\emph{Planck} provides the probability $P(H_0\vert D)$ of a cosmological background value $H_0$ given the CMB data $D$. From the local measurement one obtains the probability $P(\hat{H}_0\vert \hat{D})$ of a value $\hat{H}_0$ in our environment given the local measurement $\hat{D}$.
For simplicity, both probabilities shall be treated as Gaussians adopting the means and standard deviations quoted in Refs.~\cite{Riess:2019cxk,Aghanim:2018eyx}.
At this point it should therefore be stressed that for a more reliable likelihood analysis, a more careful implementation of the data beyond the simplified Gaussian treatment should be performed.
This is left for future work.
The analysis presented here shall only serve as an estimate of the effect of our environment on the local measurement of the Hubble constant.
Non-Gaussian features can be expected to be subdominant.

From the measurements, employing Bayes' Theorem, one may find the probability $\propto P(\hat{D}\vert H_0)$ of getting the local measurement $\hat{D}$ when given $H_0$ in the background and from that the probability $\propto P(\hat{D}\vert D)$ for a local measurement $\hat{D}$ given the CMB data $D$.
More precisely,
\begin{eqnarray}
 P(\hat{D}\vert D) & = & \int \rmd H_0 P(\hat{D}\vert H_0) P(H_0\vert D) \,, \label{eq:prob1} \\ 
 P(\hat{D}\vert H_0) & = & \int \rmd \hat{H}_0 P(\hat{D}\vert \hat{H}_0) P(\hat{H}_0\vert H_0) \,, \label{eq:prob2}
\end{eqnarray}
and $P(\hat{D}\vert \hat{H}_0) = P(\hat{H}_0\vert \hat{D}) P(\hat{D})/P(\hat{H}_0)$,
where a flat prior will be assumed for $P(\hat{H}_0)$.
Eqs.~\eqref{eq:prob1} and \eqref{eq:prob2} use that $P(\hat{D}|H_0,D)=P(\hat{D}|H_0)$ and $P(\hat{D}|\hat{H}_0,H_0)=P(\hat{D}|\hat{H}_0)$ since the measurement $\hat{D}$ is independent of $D$ given $H_0$ and independent of $H_0$ given $\hat{H}_0$, respectively.
Note that $P(\hat{D}\vert H_0)$ and $P(\hat{D}\vert D)$ can only be determined up to the factor $P(\hat{D})$.
From $P(\hat{D}\vert H_0)$ and Eq.~\eqref{eq:prob2}, one can compute the probability $P(H_0\vert \hat{D})$ for a cosmological Hubble constant $H_0$ given the local measurement up to a factor $P(H_0)^{-1}$, also assumed flat here.
$P(H_0\vert \hat{D})$ normalised at its peak is shown in Fig.~\ref{fig:results}, providing alongside a comparison to the \emph{Planck} measurement, found here to be consistent with expectations at the 95\% confidence level.
It is worth noting that the impact of environment suggests the interpretation of $P(H_0\vert \hat{D})$ as the actual local measurement of the Hubble constant of the cosmological background with mean and standard deviation $H_0 = 76.7\pm5.5$, or
more accurately
$H_0 = 74.7^{+5.8}_{-4.2}$ in terms of the median and 68\% confidence bounds.
Finally, one can assess to what extent a hypothetical local measurement $\hat{D}_{eq}$ with the same uncertainties of $\hat{D}$ but equal mean to that measured by \emph{Planck}, $\hat{H}_0=H_0$, would be more likely than the real data $\hat{D}$ obtained in Ref.~\cite{Riess:2019cxk}.
This can be estimated from the Bayes factor
\begin{equation}
 \hat{B}_{eq} = \frac{P(\hat{D}\vert D_{eq})}{P(\hat{D}\vert D)}
\end{equation}
with the unknown evidence $P(\hat{D})$ cancelling out.
The Bayes factor of this hypothetical scenario to the real measurement is given by $\hat{B}_{eq}=4$, which
while statistically preferring such a scenario
would not amount to strong evidence on the Jeffreys scale for equality over the observed case.

These considerations suggest that the measurement of the Hubble constant in the local 40~Mpc environment ($\hat{H}_0$) is marginally consistent with that inferred from the CMB by \emph{Planck} ($H_0$).

\psec{Conclusions}
%
Measurements of the current expansion rate of our Cosmos reveal a significant tension between the rate inferred from the CMB by \emph{Planck} and that found from SN~Ia separations employing absolute distance anchors from nearby Cepheids, masers, parallaxes, and ecliptic binaries.
It was argued here that the local measurement of the Hubble constant is likely impacted by a 40~Mpc underdense local region of $\denv\approx-0.5\pm0.1$ that contains the Cepheids/SN~Ia calibration sample, and hence in this context, the Hubble tension may be interpreted as a $5\sigma$ measurement of that.
This is due to the absolute distance ruler being provided by the local universe, which is not applicable to cosmological scales without appropriate transformation.
Consistent to expectations, the use of a cosmological distance ruler instead yields a measurement of $\hat{H}_0$ that is in agreement with $H_0$~\cite{Macaulay:2018fxi}.
The environmental uncertainties also suggest the reinterpretation as a measurement of the cosmological Hubble constant of $H_0=74.7^{+5.8}_{-4.2}$~km/s/Mpc.
The probability for finding the given local expansion rate for the given cosmological rate or \emph{Planck} data lies at the 95\% confidence level.
Moreover, a hypothetical data set with local expansion rate equal to that of \emph{Planck}, while statistically favoured, would not find strong preference in the Bayes factor over the actual measurement.
These results therefore suggest borderline consistency between the local and CMB measurements of the Hubble constant.

The simplified one-dimensional Gaussian treatment of the local and CMB data is a caveat to these results.
But importantly, the analysis has also been conservative with the adoption of a 40~Mpc environment, as the absolute distance anchors reside in a 10~Mpc region, where matter density fluctuations may be even more pronounced.
Due to the involved fitting procedure in the calibration of local magnitudes, the relevant scale of the environment was set to that containing all of the Cepheid/SN~Ia samples.
A more local environment set by the immediate volume of the anchors would further increase the likelihood of encountering the measured discrepancy between $H_0$ and $\hat{H}_0$ (also see Refs.~\cite{Ben-Dayan:2014swa,Yoo:2019qsl} for perturbative contributions).
Accounting for this effect would require a joint analysis of the Cepheid magnitudes in varying environments, which is beyond the scope of the current analysis and left for future work.
Such a study may also include the LIGO/Virgo gravitational wave measurement~\cite{Abbott:2017xzu} of $\hat{H}_0$ from NGC~4993, also residing in the $\sim40$~Mpc environment.
Repeated gravitational wave measurements in that volume would be expected to agree with the local expansion rate rather than with that inferred from the CMB.
The joint analysis could furthermore include the measurement of the Hubble constant by H0LiCOW quasar lensing~\cite{Wong:2019kwg}. An important point worth noting in this context is that the time delays of the images are measured with a local clock in the reference frame set by the local underdensity, and the change to the cosmic reference frame yields a cosmic Hubble parameter consistent with that inferred from the CMB (App.~\ref{sec:H0LiCOW}).
Finally, note that a consistent local underdensity has independently been confirmed by CLASSIX X-ray galaxy cluster distributions, finding $\delta_{{\rm env},0}(\lesssim40~{\rm Mpc}) = -0.5\pm0.2$~\cite{Boehringer:2019xmx}.
The local Universe is also found to be consistently underdense both at $\lesssim40$~Mpc and $\lesssim10$~Mpc from recent all-sky catalogues of galaxy groups with a range for $-\delta_{{\rm env},0}(\leq40~{\rm Mpc})$ of 0.56--0.71~\cite{Karachentsev:2018ysz}.

%
The author thanks Ruth Durrer and Steen Hansen for useful discussions and Pierre Fleury, Alex Hall, Dragan Huterer, D'Arcy Kenworthy, and Adam Riess for useful comments on the manuscript.
This work was supported by a Swiss National Science Foundation Professorship grant (No.~170547).
Please contact the author for access to research materials.

\appendix

\section{Derivations}

The following discussion provides derivations for the arguments presented around Eq.~(\ref{eq:localH0}).
For simplicity, $z\ll1$ will be assumed throughout.
Deviations of this approximation only enter through the dependence of the magnitude-redshift relation on the cosmological matter density parameter for large redshifts of the high-$z$ SN~Ia sample.
Computations can therefore easily be generalised to account for these corrections.

\subsection{Luminosity distance in a local inhomogeneity} \label{sec:app1}

\begin{figure}
\centering
 \resizebox{0.45\textwidth}{!}{
 \includegraphics{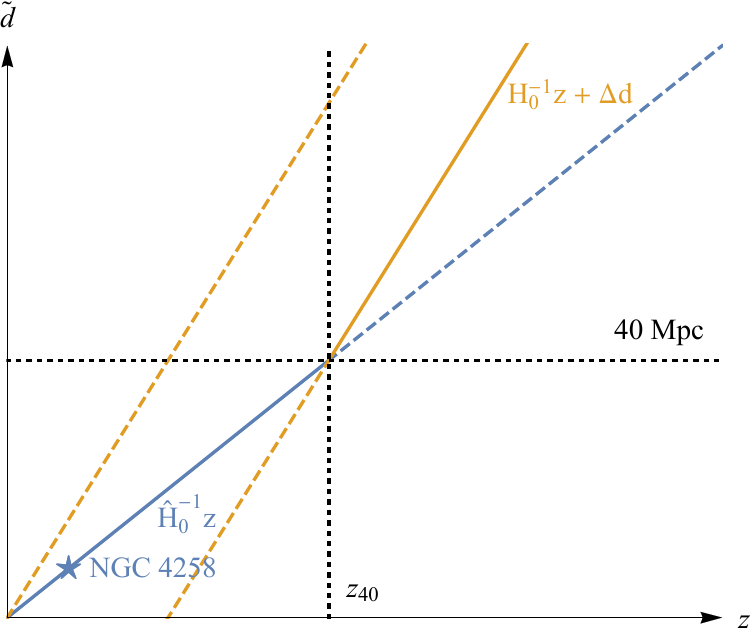}
 }
\caption{
Schematic distance-redshift relation for a uniform local inhomogeneity in a homogeneous cosmological background.
The distance scales as $\hat{d}_L=\hat{H}_0^{-1}z$ in the interior and $d_L=H_0^{-1}z+\Delta d$ in the exterior with a distance correction $\Delta d$ set at the edge of the inhomogeneity.
}
\label{fig:distanceredshift}
\end{figure}

The distance travelled by a light ray emitted by a source at redshift $z$ before observation can be illustrated by Fig.~\ref{fig:distanceredshift}.
For $z\leq z_{40} \equiv z(40~{\rm Mpc})$, it is given by $\hat{d}_L = \hat{H}_0^{-1}z$, whereas for $z>z_{40}$, $d_L = H_0^{-1}z + \Delta d$ with
\begin{equation}
 \Delta d = H_0^{-1} \left( \frac{H_0}{\hat{H}_0} - 1 \right) z_{40} \,.
\end{equation}
At large distances, $d_L \gg \Delta d$, the distance correction can be neglected.
For the distance modulus, however, it implies that whereas at low redshifts
\begin{equation}
 m - M = \tilde{\mu}_{\rm low} = 5 \log \frac{\hat{d}_L}{\hat{d}_1} + 25 = 5 \log \frac{z}{z_1} + 25
\end{equation}
with $\hat{d}_1\equiv 1~{\rm Mpc}$, at high redshifts, one instead obtains
\begin{eqnarray}
 \tilde{\mu}_{\rm high} = 5 \log \frac{d_L}{\hat{d}_1} + 25 = 5 \log \frac{z}{z_1} + 25 + \Delta M \,,
\end{eqnarray}
where a correction of the absolute magnitude was defined as
\begin{eqnarray}
 \Delta M & = & 5 \log \left[ \frac{\hat{H}_0}{H_0} + \frac{\hat{H}_0 \Delta d}{z} \right] \nonumber\\ 
 & = & 5 \log \left[ \frac{\hat{H}_0}{H_0} + \left( 1 - \frac{\hat{H}_0}{H_0}\right) \frac{z_{40}}{z} \right] \,. \label{eq:DM}
\end{eqnarray}
Thus, whereas at $z_{40}$, $\Delta M = 0$, for $z\gg z_{40}$, one obtains a shift of $\Delta M = 5 \log (\hat{H}_0/H_0)$.
This shift can be avoided if changing the measurement of distances from the local reference frame of $\hat{d}_1$ to the cosmological one of $d_1$.

\subsection{Reference frames} \label{sec:frames}

The difference between the absolute distance scales $\hat{d}_L$ and $d_L$ can be phrased in terms of two conformally related metrics, one with a scale factor $a$ that has undergone an evolution with an average cosmological matter density and one that has been governed by the local matter density, $\hat{a}$.
Let
\begin{equation}
 g_{\mu\nu} = a^2 \eta_{\mu\nu}
\end{equation}
denote the flat FLRW metric of the average universe, where $\eta_{\mu\nu}$ is the Minkowski metric.
The FLRW metric of the local universe shall instead be given by
\begin{equation}
 \hat{g}_{\mu\nu} = \hat{a}^2 \eta_{\mu\nu} \,, \label{eq:localmetric}
\end{equation}
where $\hat{a}\neq a$ for a local matter density that deviates from its cosmological average.
A metric applying to both regimes may furthermore be defined as
\begin{equation}
 \tilde{g}_{\mu\nu} = \tilde{a}^2({\bf x},\eta) \eta_{\mu\nu} \,,
\end{equation}
where $\tilde{a}=a$ for $\lvert {\bf x}-{\bf x}_0 \rvert > R$ and $\lvert {\bf x}-{\bf x}_0 \rvert \leq R$ with $R$ defining the size of the inhomogeneity.

The two homogeneous metrics can be related by a conformal factor $C$ as
\begin{equation}
 \hat{g}_{\mu\nu} = \frac{\hat{a}^2}{a^2} g_{\mu\nu} \equiv C^2 g_{\mu\nu} \,.
\end{equation}
For an emission at the cosmological time $t_{\rm em}$ in $g$, one can write
\begin{equation}
 a_{\rm em} - a_0 = \int_{t_0}^{t_{\rm em}} {\rm d}t \, \dot{a} = \int_{t_0}^{t_{\rm em}} {\rm d}t \, a H \approx a_0 H_0 (t_{\rm em}-t_0) \,.
\end{equation}
Hence, $a \approx H_0 t$.
Similarly, $\hat{a} \approx \hat{H}_0 t$ for $\hat{H}$ such that $C \approx \hat{H}_0/H_0$.

Importantly, observable quantities such as apparent magnitudes $m$ and redshifts $z$ remain invariant under the conformal transformation.
Absolute distance rulers, in contrast, are not preserved under the transformation, and as a rule of thumb, one should be careful with any quantity that carries units.

In specific, luminosity distances $d_L$ rescale as
\begin{equation}
 \hat{d}_L = C^{-1} d_L = \frac{H_0}{\hat{H}_0} d_L \,.
\end{equation}
This rescaling can also be inferred directly from
\begin{equation}
 d_L = (1+z) \int_0^z \frac{{\rm d}z}{H(z')} \approx \frac{z}{H_0}
\end{equation}
and $\hat{d}_L \approx \hat{H}_0^{-1}z$ such that
\begin{equation}
 \hat{d}_L = \frac{H_0}{\hat{H}_0} d_L \,.
\end{equation}

It is important to note that while distance moduli are invariant, $\hat{\mu}=\mu$, the adopted normalisation of distance is not (App.~\ref{sec:app1}).
Consider the distance modulus in $g$, which is given by
\begin{equation}
 m - M = \mu = 5 \log \frac{d_L}{d_1} + 25 \,,
\end{equation}
where $d_1 = d_L(z_1) \approx H_0^{-1} z_1 = 1~{\rm Mpc}$ defines the invariant redshift that corresponds to 1~Mpc in the cosmological frame.
In contrast, in $\hat{g}$, for the same redshift one finds
\begin{equation}
 \hat{\mu} = 5 \log \frac{\hat{d}_L}{\hat{d}_1} + 25 \,,
\end{equation}
where $\hat{d}_1 = \hat{d}_L(z_1) \approx \hat{H}_0^{-1} z_1 = H_0 d_1 /\hat{H}_0$.
Hence, the distance normalisation is no longer 1~Mpc.
Alternatively, one may define a new scale such that $1~\widehat{\rm Mpc}$ corresponds to $z_1$, or one may choose to absorb the change of normalisation into a shift of the reference magnitude $M$ by $\Delta M = 5 \log (\hat{H}_0/H_0)$ as in Eq.~(\ref{eq:DM}).
These issues are avoided in the distance modulus as long as luminosity distances are given in the same reference frame as the normalisation employed.
It is worth noting at this point that distances in the local universe are measured in a megaparsec scale defined by $\hat{d}_1$.

\subsection{Measurement of the Hubble constant} \label{sec:H0measurement}

The high-$z$ SN~Ia provide a fit to the magnitude-redshift relation~\cite{Riess:2009pu}
\begin{equation}
 m_x^0 = 5\log z - 5 a_x \,,
\end{equation}
whereas the low-$z$ SN~Ia are used to calibrate a conversion of $m$ into absolute distance $\hat{d}$ using the measurement of $\hat{d}_{\rm N4258}$ with its calibrated $m_{\rm N4258}$.
In the context of the conformal relation between the local and cosmological metrics, $m$ and $z$ of the high-$z$ samples can therefore be used to attribute a redshift to a distance $\hat{d}$, from which the local Hubble constant $\hat{H}_0 = z/\hat{d}$ can be inferred.
Inversely, $\hat{d}$ can be extrapolated to the high-$z$ regime, but $d$ and therefore $H_0$ remain undetermined as long as there is no absolute distance measurement provided in this reference frame (see Fig.~\ref{fig:distanceredshift}).

To perform the calculation more carefully, the magnitude-redshift relation shall first be followed from large redshifts to the edge of the local inhomogeneity at $z_{40}$, where it is matched with the magnitude-distance calibration from the low-$z$ samples.
More specifically,
\begin{eqnarray}
 a_x & = & \log z_{40} - \frac{1}{5} m_{x,40}^0 = \log \hat{H}_0 \hat{d}_{40} - \frac{1}{5} m_{x,40}^0 \nonumber\\
 & = & \log \hat{H}_0 \hat{d}_{40} - \frac{1}{5} (\hat{\mu}_{0,40} - \hat{\mu}_{0,{\rm N4258}} + m_{x,{\rm N4258}}^0) \nonumber\\
 & = & \log \hat{H}_0 - 5 - \frac{1}{5} (m_{x,{\rm N4258}}^0 - \hat{\mu}_{0,{\rm N4258}}) + \log \hat{d}_1 \,. \nonumber\\
\end{eqnarray}
Hence, one obtains Eq.~(\ref{eq:localH0}),
\begin{equation}
 \log \hat{H}_0 = \frac{1}{5} (m_{x,{\rm N4258}}^0 - \hat{\mu}_{0,{\rm N4258}}) + a_x + 5 - \log \hat{d}_1 \,. \label{eq:localH0app}
\end{equation}
Note that the distance normalisation $\hat{d}_1$ is kept here explicit.
This allows for a frame-invariant expression of the measurement of the Hubble constant.
A transformation into the cosmological reference frame gives
\begin{equation}
 \log H_0 = \frac{1}{5} (m_{x,{\rm N4258}}^0 - \mu_{0,{\rm N4258}}) + a_x + 5 - \log d_1 \,. \label{eq:cosmologicalH0}
\end{equation}
While the expression is frame invariant, the Hubble constant is a frame-dependent quantity.
The distance anchors are obtained from the local reference frame, and the Hubble constant inferred is therefore the local one.
A measurement of $H_0$ is only possible by a given absolute measurement of the cosmological ruler, or inversely, a transformation to the cosmological distance scale is only possible by additional knowledge of the cosmological Hubble constant $H_0$.
However, neither of those quantities are available from the data employed.
More explicitly, to cancel $\log d_1$ in Eq.~(\ref{eq:cosmologicalH0}) one needs to adopt the distance normalisation $d_1$ in $\mu_{0,{\rm N4258}}$.
However, since we measure $\hat{d}_{\rm N4258}$, this leads to the conversion $-\log d_{\rm N4258} = -\log(\hat{H}_0/H_0) - \log{\hat{d}_{\rm N4258}}$, which changes the measurement in Eq.~(\ref{eq:cosmologicalH0}) back into one of $\log \hat{H}_0$.

Importantly, the local environment was chosen as a uniform matter density perturbation with respect to the cosmological average density of the 40~Mpc region that contains all joint SN~Ia/Cepheid samples and distance anchors.
This is a conservative choice that was adopted to avoid any contamination of changing reference frames in the calibration of the SN~Ia magnitudes expected at the location of the distance anchors, which uses the joint SN~Ia/Cepheid data.
In this setting, distances in the Milky Way or to the Large Magellanic Cloud are measured in the same local reference frame defined by Eq.~(\ref{eq:localmetric}) with constant $\hat{a}=\hat{a}_0$.
Due to the different matter densities $\hat{a}$ has evolved differently to $a$ from early times $t_i$ when the universe was more homogeneous ($\hat{a}_i \approx a_i$).
Hence, such distances also need to be transformed for use in the cosmological reference frame.
Neglecting possible contamination in the calibration of SN~Ia magnitudes, absolute distance scales may directly be set by the distance anchors.
Realistically, their respective nearby expanding environments will deviate from that of the local 40~Mpc region.
The environment averaged over a more nearby distance may notably be expected to depart more strongly from the cosmological average, which further improves the consistency between the local and cosmological measurements of the Hubble constant.
Importantly, the expanding environments of the different distance anchors are strongly correlated due to overlapping volumes, and one may hence expect only small differences between the respective reference frames.

Finally, these results are independent of performing the calculations in the separated homogeneous frames.
The same relations are obtained if adopting the inhomogeneous metric $\tilde{g}$.
More specifically, for the high-$z$ samples one finds (Sec.~\ref{sec:app1})
\begin{eqnarray}
 \tilde{\mu}_0 - \tilde{\mu}_{0,40} & = & m_x^0 - m_{x,40}^0 + \Delta M \nonumber\\
 5 \log \frac{d_L}{d_{40}} & = & 5 \log \frac{z}{z_{40}} + \Delta M \,. 
\end{eqnarray}
This implies that
\begin{eqnarray}
 & & \frac{1}{5} (m_{x,{\rm N4258}}^0 - \hat{\mu}_{0,{\rm N4258}}) = \frac{1}{5} (m_{x,40}^0 - \tilde{\mu}_{0,40}) \nonumber\\
 & & \quad = \frac{1}{5} (m_x^0 - \tilde{\mu}_0 + \Delta M ) \nonumber\\
 & & \quad = \frac{1}{5} (m_x^0 + \Delta M) - 5 - \log d_L + \log \hat{d}_1 \nonumber\\
 & & \quad =\log z - a_x + \frac{1}{5}\Delta M - 5 - \log (H_0^{-1}z + \Delta d) + \log \hat{d}_1 \nonumber\\
 & & \quad = \log \hat{H}_0 - a_x - 5 + \log \hat{d}_1 \,, \nonumber \\
\end{eqnarray}
where the last equality follows from Eq.~(\ref{eq:DM}).
Hence, one recovers Eq.~(\ref{eq:localH0app}).

\subsection{Comment on H0LiCOW measurement} \label{sec:H0LiCOW}

Measuring the time delay between two images of gravitationally lensed quasars, H0LiCOW~\cite{Wong:2019kwg} reported a 2.4\% precision measurement of a high value of the Hubble constant consistent with that from the local distance ladder.
This delay is specified by $\Delta t_{ij} = D_{\Delta t}[\ldots]$, where the bracket is a function of angular positions of the images $i$ and $j$ and the source, $D_{\Delta t} = (1+z_d) D_d D_s/D_{ds} \propto H_0^{-1}$ with angular diameter distances $D$ and indices $d/s$ referring to the lens and source.

Importantly, this time interval is measured with a local clock in the local reference frame, $\Delta t_{ij} = \hat{a}_0 \Delta \eta_{ij}$ where $\eta$ is the conformal time and the normalisation $\hat{a}_0=1$ has implicitly been adopted in the measurement.
This needs to be transformed into the cosmic reference frame for inference of the cosmic Hubble parameter, where the conformal factor changing the reference frame causes a slow down or speed up between the local and cosmic clocks.
More specifically,
\begin{eqnarray}
 \Delta \eta_{ij} & \propto & \frac{1}{\hat{a}_0} \frac{a_0}{a} \frac{D_d D_s}{D_{ds}} = \frac{a_0}{\hat{a}_0} D_{\Delta t} = C^{-1} D_{\Delta t} \nonumber\\
 & \propto & \frac{H_0}{\hat{H}_0} H_0^{-1} = \hat{H}_0^{-1} \,.
\end{eqnarray}
Hence, H0LiCOW measured the local $\hat{H}_0$ rather than the cosmic $H_0$.
Applying the conformal factor $C$ for the change to the cosmic reference frame, one obtains
\begin{equation}
 C^{-1}\Delta t_{ij}^{-1} |_{\hat{a}_0\equiv1} = C^{-1} \Delta \eta_{ij}^{-1} \propto \frac{H_0}{\hat{H}_0} \hat{H}_0 = H_0 \,.
\end{equation}
With the value of the conformal factor inferred from the local distance ladder the H0LiCOW measurement $\hat{H}_0=73.3^{+1.7}_{-1.8}$~km/s/Mpc therefore yields a cosmic $H_0$ consistent with that of \emph{Planck}.

\bibliographystyle{arxiv_physrev_mod}
\bibliography{hubbleconstant}

\end{document}